\documentclass{article}

\PassOptionsToPackage{numbers, compress}{natbib}

\usepackage[preprint]{neurips_2025}

\usepackage[utf8]{inputenc} 
\usepackage[T1]{fontenc}    
\usepackage{hyperref}       
\usepackage{url}            
\usepackage{booktabs}       
\usepackage{amsfonts}       
\usepackage{nicefrac}       
\usepackage{microtype}      
\usepackage{graphicx}       
\usepackage{subcaption}     
\usepackage{multirow}
\usepackage[table]{xcolor}

\usepackage{algorithm}
\usepackage{algpseudocode}
\usepackage{amsmath}
\usepackage{amssymb}

\usepackage{booktabs}
\usepackage{pifont} 
\newcommand{\cmark}{\checkmark}
\newcommand{\xmark}{\ding{55}}

\title{HALO: Hierarchical Autonomous Logic-Oriented Orchestration for Multi-Agent LLM Systems}
\label{title}

\author{
    Zhipeng Hou\thanks{First author and corresponding author, correspondence to \href{mailto:japhonehou@gmail.com}{\texttt{japhonehou@gmail.com}}} \\
    Nanjing University of Posts and Telecommunications\\
    \href{mailto:japhonehou@gmail.com}{\texttt{japhonehou@gmail.com}} \\
    \And
    Junyi Tang\\
    Nanjing University of Posts and Telecommunications\\
    \href{mailto:b23012214@njupt.edu.cn}{\texttt{b23012214@njupt.edu.cn}} \\
    \And
    Yipeng Wang\\
    Chongqing University\\
    \href{mailto:yipengxw@gmail.com}{\texttt{yipengxw@gmail.com}} \\
}

\begin{document}

\maketitle

\begin{abstract}
    \label{abstract}

    Recent advancements in Multi-Agent Systems (MAS) powered by Large Language Models (LLMs) have demonstrated tremendous potential in diverse task scenarios. Nonetheless, existing agentic systems typically rely on predefined agent-role design spaces and static communication structures, limiting their adaptability as well as flexibility in complex interaction environments and leading to subpar performance on highly specialized and expert-level tasks.
    To address these issues, we introduce \textbf{HALO}, a multi-agent collaboration framework based on a hierarchical reasoning architecture. Specifically, we incorporate a high-level planning agent for task decomposition, mid-level role-design agents for subtask-specific agent instantiation, and low-level inference agents for subtask execution. Particularly, subtask execution is reformulated as a structured workflow search problem, where Monte Carlo Tree Search (MCTS) systematically explores the agentic action space to construct optimal reasoning trajectories. Additionally, as the majority of users lack expertise in prompt engineering, we leverage an Adaptive Prompt Refinement module to transform raw queries into task-specific prompts.
    Empirical evaluations on Code Generation (HumanEval), General Reasoning (MMLU), and Arithmetic Reasoning (MATH) benchmark datasets highlight the effectiveness of HALO, yielding a \textbf{14.4\%} average improvement over state-of-the-art baselines. Notably, HALO achieves up to \textbf{13.3\%} performance gain on the Moral Scenarios subject in the MMLU benchmark and up to \textbf{19.6\%} performance gain on the Algebra subarea in the MATH benchmark, indicating its advanced proficiency in tackling highly specialized and expert-level tasks.
    The code repository is available at \url{https://github.com/23japhone/HALO}.
\end{abstract}

\section{Introduction}
\label{sec_1_introduction}

Large Language Models (LLMs), such as OpenAI o3~\cite{el2025competitive} and Deepseek R1~\cite{deepseekai2025deepseekr1incentivizingreasoningcapability}, have demonstrated remarkable capabilities in both language understanding and reasoning. These advancements unlock tremendous potential for LLM-based multi-agent systems to address a broad spectrum of downstream tasks, including code generation~\cite{zhong2024debug,shinn2023reflexion}, mobile device control~\cite{wang2024mobile,li2024appagent}, video gaming~\cite{wang2023voyager} and open-domain question answering~\cite{li2022ai,wang2024chain}. Consequently, the formulation of effective agentic systems is crucial to fully harnessing the capabilities of LLMs across diverse downstream applications.


However, existing agentic systems often struggle to maintain robust performance in complex interaction environments and expert-level tasks. This limitation arises from predefined agent-role design spaces~\cite{wang2024mobile,zeng2024perceive} and static communication workflows~\cite{yao2023react,li2023camel}, which heavily depend on expert insight and manually-designed policies. Meanwhile, as the majority of users lack expertise in prompt engineering, poorly formulated queries have greatly hindered the comprehension of agents, ultimately leading to inefficient task execution. Together, these challenges underscore two fundamental problems: (1) how can agentic systems self-organize and coordinate in unfamiliar environments with minimal manual intervention; and (2) how can user queries be refined to improve the overall efficiency and effectiveness of multi-agent collaboration?

\begin{figure}[t]
    \centering
    \includegraphics[width=\linewidth]{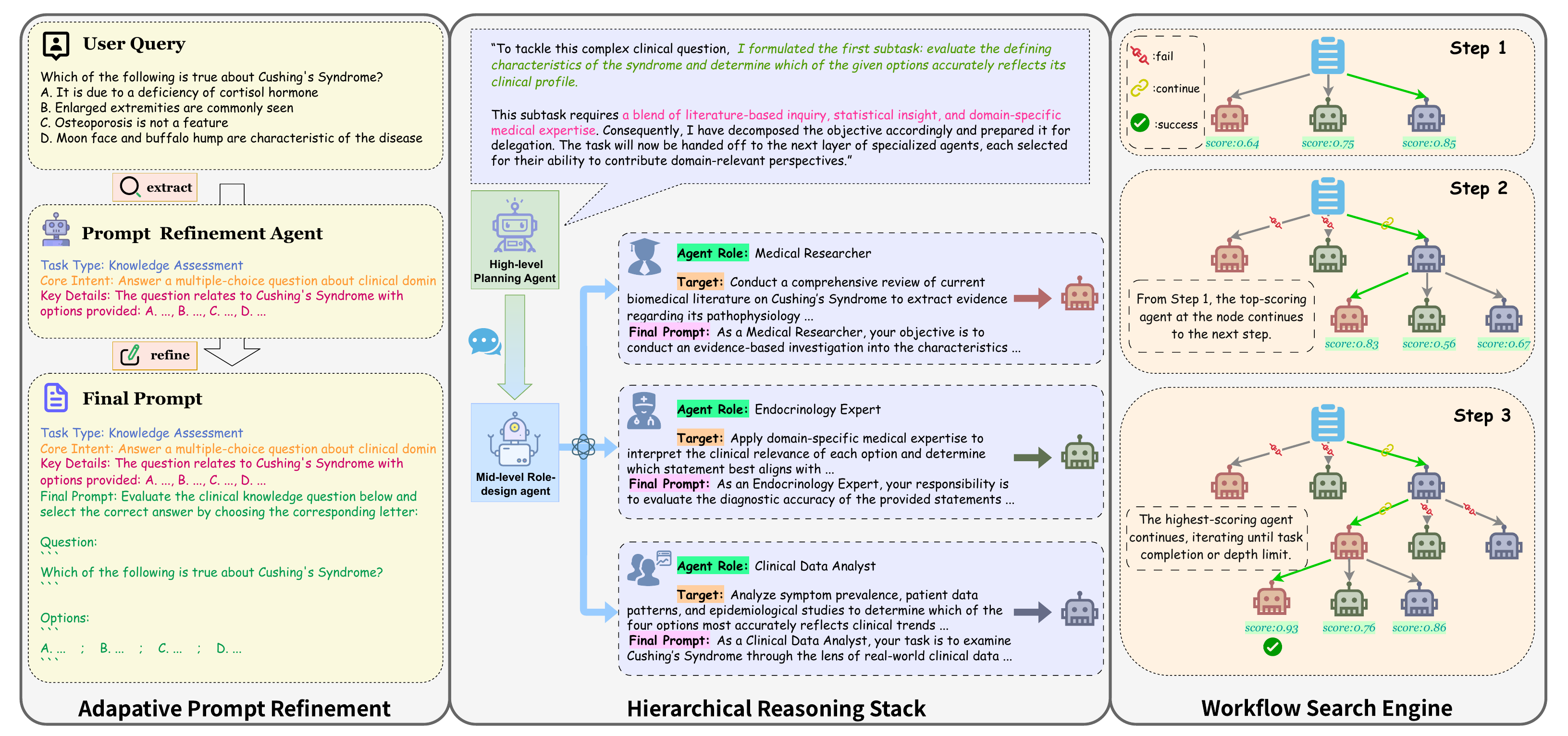}
    \caption{
        The overview of HALO framework. HALO consists of three modules: (1) Adaptive Prompt Refinement (Section~\ref{sec_3.2_prompt_refinement}), where user queries are refined into high-quality and LLM-comprehensible prompts; (2) Hierarchical Reasoning Stack (Section~\ref{sec_3.3_hierarchical_reasoning}), which is responsible for task decomposition, role instantiation, and subtask execution; and (3) Workflow Search Engine (Section~\ref{sec_3.4_workflow_search}), which explores multi-agent collaboration and constructs optimal workflows. Green paths denote optimal reasoning trajectories, while red paths are pruned during search.
    }
    \label{fig_1_overview}
\end{figure}

In response, we introduce \textbf{HALO}, a \textbf{H}ierarchical \textbf{A}utonomous \textbf{L}ogic-Oriented \textbf{O}rchestration framework focused on addressing complex interaction environments and expert-domain reasoning tasks. To this end, HALO incorporates an extensible agent-role instantiation mechanism and a dynamic communication architecture to replace the rigidity of predefined role spaces and static workflows. Additionally, HALO employs a prompt engineering module for user query refinement. The overview of HALO is illustrated in Figure~\ref{fig_1_overview}.

Specifically, HALO functions in a three-stage paradigm.
The first stage is \textbf{Adaptive Prompt Refinement} (detail in Section~\ref{sec_3.2_prompt_refinement}), where the raw user query is transformed into a high-quality and LLM-comprehensible prompt. This module comprises four collaborative agents responsible for different phase of prompt refinement, ranging from query parsing to final synthesis.
The second stage is \textbf{Hierarchical Reasoning Stack} (detail in Section~\ref{sec_3.3_hierarchical_reasoning}), which employs a three-tier multi-agent collaboration architecture. At the top layer, a high-level planning agent decomposes the overall task into a sequence of subtasks. At the middle layer, mid-level role-design agents dynamically instantiate specialized agents tailored to the requirement of each subtask. At the bottom layer, low-level inference agents are responsible for executing each subtask through cooperation mechanisms.
The third stage is \textbf{Workflow Search Engine} (detail in Section~\ref{sec_3.4_workflow_search}), which explores low-level inference agents collaboration and constructs optimal workflows. We reformulate subtask execution as a workflow search process, where Monte Carlo Tree Search (MCTS)~\cite{browne2012survey} guides the exploration over action spaces. Each node in the search tree corresponds to an agent-generated response or an intermediate reasoning step, while edges denote possible transitions between reasoning states. Together, these nodes and edges form a path that represents a candidate reasoning trajectory for each subtask.

Extensive experiments (detail in Section~\ref{sec_4_2_main_results}) demonstrate that HALO outperforms strong baselines across Code Generation (HumanEval)~\cite{chen2021evaluating}, General Reasoning (MMLU)~\cite{hendrycks2020measuring}, and Arithmetic Reasoning (MATH)~\cite{hendrycks2021measuring} benchmarks, yielding a 14.4\% average improvement. Notably, HALO achieves up to \textbf{13.3\%} performance gain on the Moral Scenarios subject in MMLU and up to \textbf{19.6\%} performance gain on the Algebra subarea in MATH, underscoring its advanced proficiency in highly specialized and expert-level tasks.

The key contributions of this work are as follows:
\begin{itemize}
    \item We introduce a novel framework named HALO for task-oriented agent collaboration in three stages, marking a significant advancement beyond the limitations of expert insight and manually-designed policies.
    \item Experiments across three diverse tasks show that our method outperforms state-of-the-art baselines, confirming the effectiveness and adaptability of HALO across complex interaction environments and expert-domain reasoning tasks.
\end{itemize}

\section{Related work}
\label{sec_2_related_work}

\subsection{Prompt optimization}
\label{sec_prompt_optimization}
Prompt design underpins how LLM agents interpret instructions and coordinate actions. Recent work automates this process to reduce manual intervention and improve modularity. Promptor~\cite{shen2023promptorconversationalautonomousprompt} decomposes goals into structured, role-specific prompts, enabling agents to focus on individual tasks. CAMEL~\cite{li2023camel} integrates a task-planner agent that generates role prompts to guide multi-agent interactions through instruction-driven simulation.
While these methods improve modularity and task decomposition, current agentic systems still depend on handcrafted prompt templates or task-specific engineering, which limits generalization across domains. Recent frameworks like ComfyAgent~\cite{xue2024comfybenchbenchmarkingllmbasedagents} begin to incorporate prompt generation as part of evolving workflows. Nonetheless, the challenge of designing prompts that are both reusable and responsive to runtime context changes remains largely open, especially in scenarios where agent roles or task objectives are not predefined.

\subsection{Role design in LLM-based architecture}
\label{sec_role_design_in_LLM_based_architecture}
Role design has become a central theme in recent multi-agent architectures. Manual role assignment is a common strategy for organizing agent behavior. MetaGPT~\cite{hong2023metagpt}, for instance, aligns agents with real-world job titles and governs their behavior through standard operating procedures. This static role configuration offers clarity and reproducibility, especially in well-defined domains.
However, fixed roles often struggle in open-ended or dynamic settings where agent responsibilities must evolve in response to task changes.
To address this, newer frameworks introduce role generation mechanisms that allow agents to create, adapt, or inherit roles dynamically. TPTU~\cite{ruan2023tptu} and DyLAN~\cite{liu2024dynamic} adopt hierarchical or layered structures where agents can delegate, transform, or refine responsibilities as collaboration progresses. 
These approaches mark a shift toward greater flexibility, although existing agentic systems still operate within partially predefined boundaries.

\subsection{Cooperation optimization strategies}
\label{sec:cooperation_optimization}
As the scale and complexity of LLM-based multi-agent systems grow, optimizing inter-agent cooperation becomes increasingly critical. Broadly, current systems adopt either centralized or decentralized paradigms.
Centralized frameworks, such as AgentLaboratory~\cite{schmidgall2025agent}, ScoreFlow~\cite{wang2025scoreflow}, and WORKFLOW-LLM~\cite{fan2024workflowllm}, rely on controller agents to manage scheduling, communication, and output aggregation.
While these architectures offer clear oversight, they can suffer from scalability issues and become bottlenecks in real-time or asynchronous settings.
In contrast, decentralized approaches seek to distribute control and decision-making across agents.
Some utilize peer-to-peer communication or identity-based protocols~\cite{liu2025advances}, while others apply game-theoretic negotiation~\cite{hua2024gametheoreticllmagentworkflow}.
These designs offer greater autonomy and robustness in unstructured environments.
Recently, several frameworks reframe cooperation as a structured search process. AFlow~\cite{zhang2025aflowautomatingagenticworkflow}, for instance, explores layered task workflows through dynamic graph traversal rather than relying on fixed execution paths. Such methods enable agents to adaptively construct interaction plans based on task-specific signals or peer responses, highlighting a promising direction for building more context-sensitive agent teams.
Notably, some researchs apply reinforcement learning to refine cooperation strategies~\cite{yu2024fincon, bucsoniu2010multi, arel2010reinforcement, zhang2021multi, kapetanakis2003reinforcement}.

\begin{table}[t]
    \centering
    \caption{Comparative analysis of HALO and existing representative LLM-based multi-agent frameworks across five key dimensions.}
    \vspace{\baselineskip}
    \label{tab_1_diverse_architecture}
    \resizebox{\textwidth}{!}{
        \begin{tabular}{llccccc}
            \toprule
            \textbf{Method} & \textbf{Structure} & \textbf{Multi-Role} & \textbf{Role Assignment} & \textbf{Dynamic Structure} & \textbf{Team Optimization} \\
            \midrule
            ScoreFlow~\cite{wang2025scoreflow} & Central + Score-based & \xmark & \xmark & \xmark & \cmark \\
            CAMEL~\cite{li2023camel} & Feedback Triplet & Manual & Manual & \xmark & \xmark \\
            AgentVerse~\cite{chen2023agentversefacilitatingmultiagentcollaboration} & Hierarchical Tree & \cmark & Generated & \xmark & \xmark \\
            MetaGPT~\cite{hong2023metagpt} & SOP-guided Workflow & \cmark & Manual & \xmark & \xmark \\
            DyLAN~\cite{liu2024dynamic} & DAG + Feedback Loop & \cmark & Manual + Gen & \cmark & \cmark \\
            ComfyAgent~\cite{xue2024comfybenchbenchmarkingllmbasedagents} & Modular Workflow & \cmark & Generated & \cmark & \xmark \\
            AgentLaboratory~\cite{schmidgall2025agent} & Centralized Scheduler & \cmark & Manual & \xmark & \cmark \\
            TPTU~\cite{ruan2023tptu} & Hierarchical Planner & \cmark & Manual + Gen & \cmark & \xmark \\
            AFlow~\cite{zhang2025aflowautomatingagenticworkflow} & Multi-layer DAG & \cmark & Manual + Gen & \cmark & \cmark \\
            WORKFLOW-LLM~\cite{fan2024workflowllm} & Centralized Workflow & \cmark & Manual & \xmark & \cmark \\
            Game-theoretic Workflow~\cite{hua2024gametheoreticllmagentworkflow} & Decentralized Negotiation & \cmark & Adaptive & \cmark & \cmark \\
            \cellcolor{gray!20}\textbf{HALO (Ours)} & \cellcolor{gray!20}\textbf{Hierarchical Structure + MCTS} & \cellcolor{gray!20}\cmark & \cellcolor{gray!20}Adaptive & \cellcolor{gray!20}\cmark & \cellcolor{gray!20}\cmark \\
            \bottomrule
        \end{tabular}
    }
\end{table}

A comparative summary of these cooperation frameworks across five dimensions is provided in Table~\ref{tab_1_diverse_architecture}. In contrast, HALO introduces a hierarchical reasoning architecture and reformulates subtask execution as a structured reasoning search. This design offers a balanced pathway between modular coordination and dynamic adaptability, particularly suited for complex interaction environments.

\section{HALO framework}
\label{sec_3_HALO_framework}

\subsection{Problem formulation}
\label{sec_3.1_problem_formulation}

In the proposed workflow search process, a workflow $\mathcal{W}$ is composed of a sequence of subtasks $\{T_1, T_2, \dots, T_K\}$, where each subtask $T_k$ is handled by a series of role-specialized LLM-based agents $\mathcal{A}_k = \{a_k^{(1)}, a_k^{(2)}, \dots\}$. We reformulate subtask execution as the workflow search process over the workflow space $\mathcal{S}$, where each candidate workflow $\mathcal{W} \in \mathcal{S}$ represents a unique instantiation of subtasks and multi-agent interactions. Given a user query $\mathcal{Q}$ expressed in natural language, the objective is to construct an optimal reasoning workflow $\mathcal{W}^*$ that generates an expected answer $\hat{Y}$ aligned with the user query $\mathcal{Q}$. The optimization problem of HALO is defined as follows:
\begin{equation}
    \mathcal{W}^* = \arg\max_{\mathcal{W} \in \mathcal{S}} \ \mathrm{Value}(\mathcal{Q}, \mathcal{W})
\end{equation}
where $\mathrm{Value}(\mathcal{Q}, \mathcal{W})$ evaluates the effectiveness of workflow $\mathcal{W}$ in addressing the user query $\mathcal{Q}$.

The overall framework of HALO is illustrated in Figure~\ref{fig_1_overview} and the algorithm is presented in Algorithm~\ref{alg:halo}. Next, we elaborate on each component of the three-stage paradigm.

\begin{algorithm}[t]
    \caption{Algorithm of HALO}
    \label{alg:halo}
    \textbf{Input:} User query $\mathcal{Q}$, prompt refinement agents $\mathcal{P}_1 \sim \mathcal{P}_4$, planning agent $\mathcal{A}_{\text{plan}}$, role-design agent $\mathcal{A}_{\text{role}}$, inference agent pool $\mathcal{A}_k$, subtask budget $K$ \\
    \textbf{Output:} Final answer $\hat{Y}$
    
    \begin{algorithmic}[1]
        \State $\mathcal{F} \gets \mathcal{P}_1(\mathcal{Q})$ \Comment{Parse task type $\mathcal{T}$, intent $\mathcal{I}$, and details $\mathcal{D}$}
        \State $\mathcal{Q}_0 \gets \mathcal{P}_2(\mathcal{Q}, \mathcal{F})$ \Comment{Build initial prompt template}
        \State $\mathcal{Q}_{\text{opt}} \gets \mathcal{P}_3(\mathcal{Q}_0, \mathcal{F})$ \Comment{Optimize with prompting strategies}
        \State $\mathcal{Q}^* \gets \mathcal{P}_4(\mathcal{Q}_{\text{opt}}, \mathcal{F})$ \Comment{Synthesize final prompt}
        
        \State $H_0 \gets \emptyset$ \Comment{Initialize subtask execution history}
        \For{$k = 1$ to $K$}
            \State $T_k \gets \mathcal{A}_{\text{plan}}(\mathcal{Q}^*, \mathcal{F}, H_{k-1})$ \Comment{Decompose query into subtask}
            \If{$T_k = \texttt{stop}$}
                \State \textbf{break} \Comment{Terminate if planning agent signals task completion}
            \EndIf        
            \State $\mathcal{A}_k \gets \mathcal{A}_{\text{role}}(T_k, \mathcal{Q}^*, \mathcal{F})$ \Comment{Generate role-specific agents}
            \For{each agent $a_k^{(i)} \in \mathcal{A}_k$}
                \State $y_k^{(i)} \gets a_k^{(i)}(T_k, \mathcal{Q}^*, \mathcal{F})$ \Comment{Generate intermediate output by executing subtask $T_k$}
                \State $\ell_k^{(i)}, v_k^{(i)} \gets \text{Evaluate}(y_k^{(i)})$ \Comment{Evaluate status label and quality score for output}
                \State $\text{MCTS\_Backpropagate}(a_k^{(i)}, \ell_k^{(i)}, v_k^{(i)})$ \Comment{Propagate execution feedback}
            \EndFor
            \State $H_k \gets H_{k-1} \cup \{T_k, \hat{Y}_k\}$ \Comment{Update subtask execution history}
            \If{early-stop($H_k$) is True}
                \State \textbf{break} \Comment{Terminate if early-stopping mechanism is triggered}
            \EndIf
        \EndFor
        \State $\hat{Y} \gets \text{Aggregate}(\{y_k^{(i)}\})$ \Comment{Select final answer}
        \State \Return $\hat{Y}$
    \end{algorithmic}
\end{algorithm}

\subsection{Adaptive Prompt Refinement}
\label{sec_3.2_prompt_refinement}

As the majority of users lack expertise in prompt engineering, user queries $\mathcal{Q}$ are often loosely structured or ambiguous in intents. To address this limitation and enhance the reliability of downstream reasoning, we propose an Adaptive Prompt Refinement module at the beginning of the HALO framework, which refines the raw user query $\mathcal{Q}$ into a structured and LLM-comprehensible prompt $\mathcal{Q}^*$. This module comprises four collaborative agents, each formalized as a functional mapping, denoted by $\mathcal{P}_1$ to $\mathcal{P}_4$. These agents are responsible for different stages of prompt refinement, ranging from query parsing to final synthesis (detail in Appendix~\ref{appendix_refinement_prompt}). The process is formulated as follows.

The process begins with the Task Parser Agent $\mathcal{P}_1$, which semantically analyzes the original query $\mathcal{Q}$ to extract three essential components: the task type $\mathcal{T}$, the core intent $\mathcal{I}$, and the key details $\mathcal{D}$. These components are then assembled into a structured triplet:
\begin{equation}
    \mathcal{F} = \mathcal{P}_1(\mathcal{Q}) = (\mathcal{T}, \mathcal{I}, \mathcal{D})
\end{equation}
This structured task representation $\mathcal{F}$ serves as a global semantic context throughout the entire reasoning workflow.

Based on this structured task representation $\mathcal{F}$ and raw user query $\mathcal{Q}$, the Prompt Template Agent $\mathcal{P}_2$ constructs an initial prompt frame $\mathcal{Q}_0$ that includes a reformulated task description, clear reasoning objectives, bounded input conditions, and the explicit output format:
\begin{equation}
    \mathcal{Q}_0 = \mathcal{P}_2(\mathcal{Q}, \mathcal{F})
\end{equation}

To further refine the initial prompt template, the Prompt Optimization Agent $\mathcal{P}_3$ incorporates slow-thinking prompting strategies~\cite{wei2022chain,wang2022self,zhou2022least,yao2023tree,kojima2022large} as well as tool calling~\cite{hou2025modelcontextprotocolmcp} instructions, generating an optimized prompt $\mathcal{Q}_{\text{opt}}$:
\begin{equation}
    \mathcal{Q}_{\text{opt}} = \mathcal{P}_3(\mathcal{Q}_0, \mathcal{F})
\end{equation}

Finally, the Prompt Generator Agent $\mathcal{P}_4$ synthesizes the optimized structure into the final refined prompt $\mathcal{Q}^*$, which serves as the entry point for downstream multi-agent reasoning:
\begin{equation}
    \mathcal{Q}^* = \mathcal{P}_4(\mathcal{Q}_{\text{opt}}, \mathcal{F})
\end{equation}

\subsection{Hierarchical Reasoning Stack}
\label{sec_3.3_hierarchical_reasoning}

Following the refined prompt $\mathcal{Q}^*$, HALO proceeds to the Hierarchical Reasoning Stack module, which comprises three collaborative layers, including the high-level planning agent $\mathcal{A}_{\text{plan}}$ for task decomposition, the mid-level role-design agents $\mathcal{A}_{\text{role}}$ for dynamic agent instantiation, and the low-level inference agents $\mathcal{A}_k = \{a_k^{(1)}, a_k^{(2)}, \dots\}$ for subtask execution.

At the top layer, the high-level planning agent $\mathcal{A}_{\text{plan}}$ receives the refined prompt $\mathcal{Q}^*$ and the global structured task representation $\mathcal{F}$. Based on this information, it decomposes the overall task into a sequence of subtasks $\{T_1, T_2, \dots, T_K\}$ (detail in Appendix~\ref{appendix_high_workflow}). Instead of performing full decomposition in advance, $\mathcal{A}_{\text{plan}}$ adopts a step-wise strategy, generating one subtask at a time and iteratively updating its decomposition policy based on the execution history of preceding subtasks:
\begin{equation}
    T_k = \mathcal{A}_{\text{plan}}(\mathcal{Q}^*, \mathcal{F}, H_{k-1})
\end{equation}
where $H_{k-1}$ denotes the execution history of preceding subtasks for the subtask $T_k$.

Each generated subtask $T_k$ is delegated to the mid-level role-design agents $\mathcal{A}_{\text{role}}$, which instantiate a set of specialized LLM-based agents $\mathcal{A}_k = \{a_k^{(1)}, a_k^{(2)}, \dots\}$ to execute the subtask $T_k$. This instantiation process is jointly guided by the semantics of the subtask $T_k$, the refined prompt $\mathcal{Q}^*$, and the global structured task representation $\mathcal{F}$, ensuring that each generated agent $a_k^{(i)}$ is well aligned with the subtask $T_k$ requirements. For each generated agent $a_k^{(i)}$, a role-specific system prompt $\rho_k^{(i)}$ is assigned to govern its behavior within the subtask $T_k$ (detail in Appendix~\ref{appendix_mid_level_prompt}):
\begin{equation}
    a_k^{(i)} = \mathcal{A}_{\text{role}}(T_k, \mathcal{Q}^*, \mathcal{F}) \Rightarrow \rho_k^{(i)}
\end{equation}

For each subtask $T_k$, the low-level inference agents $\mathcal{A}_k = \{a_k^{(1)}, a_k^{(2)}, \dots\}$ receives the subtask $T_k$, the refined prompt $\mathcal{Q}^*$, and the global structured task representation $\mathcal{F}$. They are subsequently engaged in collaborative reasoning, resulting in a set of intermediate outputs:
\begin{equation}
    \mathcal{Y}_k = \{y_k^{(1)}, y_k^{(2)}, \dots\} = \mathcal{A}_k(T_k, \mathcal{Q}^*, \mathcal{F})
\end{equation}

To further improve efficiency and avoid unnecessary subtask generation, HALO integrates an early-stopping mechanism within the high-level planning agent $\mathcal{A}_{\text{plan}}$. This design is inspired by the Byzantine Consensus theory~\cite{castro1999practical}, which states that at least $3p + 1$ agents are required to tolerate $p$ faulty agents in a single round of communication. Following this principle, HALO terminates the reasoning process if at least $66\%$ of the completed subtasks in the execution history $H_{k}$ yield a consistent answer $\hat{Y}$. Additionally, the reasoning process will also be terminated when the maximum number of permitted subtasks has been reached.

\subsection{Workflow Search Engine}
\label{sec_3.4_workflow_search}

\begin{figure}[t]
    \centering
    \includegraphics[width=\linewidth]{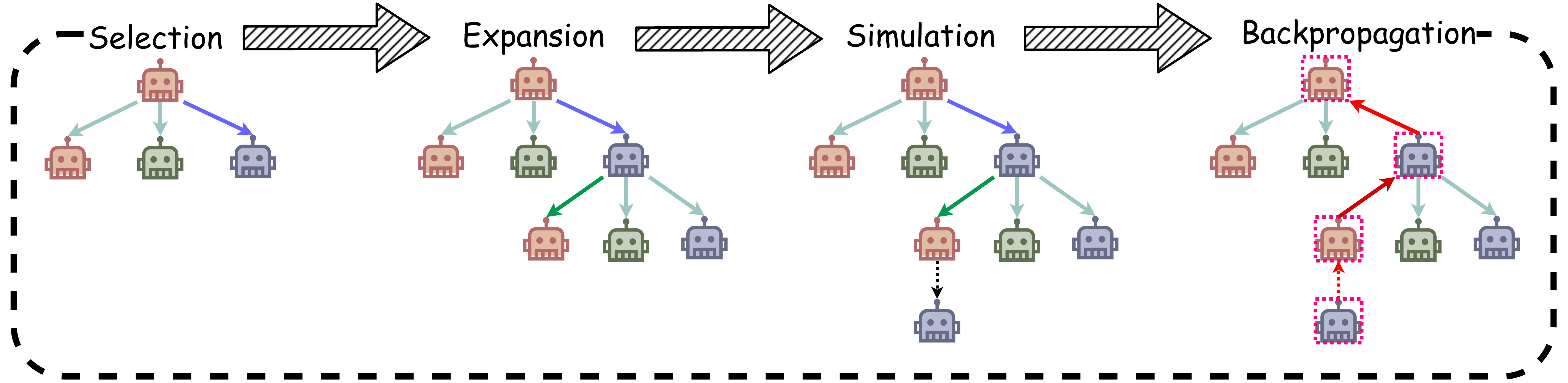}
    \caption{
        The illustration of how Monte Carlo Tree Search (MCTS) guides multi-agent reasoning through selection, expansion, simulation, and backpropagation stages. Each node represents an agent and edge transitions are guided by execution outcomes as well as evaluation feedback.
    } 
    \label{fig_2_mcts}
\end{figure}

To dynamically explore multi-agent collaboration and adaptively construct the optimal workflow, HALO leverages a Workflow Search Engine module based on Monte Carlo Tree Search (MCTS)~\cite{browne2012survey}. The workflow search process formalizes multi-agent inference as a tree-structured action space, where each node represents a role-specific agent $a_k^{(i)}$ executing a subtask $T_k$ and each edge denotes communication transitions between agents. Together, these nodes and edges form a path that corresponds to a candidate reasoning trajectory for each subtask $T_k$. During each search iteration, HALO follows the standard four-stage optimized MCTS paradigm (Selection, Expansion, Simulation, and Backpropagation), as illustrated in Figure~\ref{fig_2_mcts}.

\paragraph{Selection.}
HALO recursively selects the best agent $a_k^{(i)}$ with the UCT algorithm~\cite{kocsis2006bandit} and adds it to the trajectory:
\begin{equation}
\mathrm{UCT}(a_k^{(i)}) = \frac{v_k^{(i)}}{n_k^{(i)}} + \alpha \sqrt{\frac{\log N}{n_k^{(i)}}}
\end{equation}
where $v_k^{(i)}$ is the score value of agent $a_k^{(i)}$, $n_k^{(i)}$ is the number of visits to the agent $a_k^{(i)}$, $N$ is the number of visits to the parent agent $a_k^{(i-1)}$, and $\alpha$ is an exploration coefficient.

\paragraph{Expansion.}
Given a selected agent $a_k^{(i)}$ from the selection phase, if it contains untried actions, HALO expands the search tree by instantiating a new role-specific agent $a_k^{(i+1)}$ as its child.

\paragraph{Simulation.}
The agent $a_k^{(i+1)}$ initiates a simulated reasoning trajectory for subtask $T_k$, where a sequence of additional agents $\tilde{a}_k^{(i+2)}, \tilde{a}_k^{(i+3)}, \dots$ are engaged to emulate hypothetical future steps along the reasoning path. Each simulated agent $\tilde{a}_k^{(j)}$ generates an intermediate output $\tilde{y}_k^{(j)}$:
\begin{equation}
    \tilde{y}_k^{(j)} = \tilde{a}_k^{(j)}(T_k, \mathcal{Q}^*, \mathcal{F})
\end{equation}
The output $\tilde{y}_k^{(j)}$ is evaluated by two auxiliary agents. The judging agent assigns a status label $\tilde{\ell}_k^{(j)} \in \{success, fail, continue\}$, indicating whether the subtask is completed. The scoring agent computes a quality score value $\tilde{v}_k^{(j)} \in [0, 1]$ to reflect the effectiveness of the generated output (detail in Appendix~\ref{appendix_high_workflow}).

\paragraph{Backpropagation.}
Inspired by the value aggregation strategy in CoAT~\cite{pan2025coat}, HALO updates the evaluation scores of all traversed nodes along the search path by incorporating simulation feedback. To enhance sensitivity to task completion status, we additionally introduce a reward signal adjustment mechanism based on the judgment outcome.

Let $\lambda(\tilde{\ell}_k^{(j)})$ be the impact factor associated with the status label $\tilde{\ell}_k^{(j)}$, where $\lambda$ reflects the reward or penalty of a simulation result. The updated value of a node $a_k^{(i)}$ is computed as follows:
\begin{equation}
v_k^{(i)*} = \frac{v_k^{(i)} \cdot n_k^{(i)} + \lambda(\tilde{\ell}_k^{(j^\dagger)}) + \sum\limits_{\tilde{a}_k^{(j)} \in \mathrm{Child}(a_k^{(i)})} \tilde{v}_k^{(j)}}{n_k^{(i)} + |\mathrm{Child}(a_k^{(i)})|}
\end{equation}
where $\tilde{\ell}_k^{(j^\dagger)}$ denotes the terminal status label of the simulated leaf node agent $\tilde{a}_k^{(j^\dagger)}$ along the reasoning trajectory and $\mathrm{Child}(a_k^{(i)})$ refers to the set of its simulated children.

\section{Experiments}
\label{sec_4_experiment}

\begin{table}[t]
    \centering
    \caption{
        Performance comparison of HALO against competitive baselines across three benchmarks. Metrics include $pass@1$ (\%) for HumanEval, $accuracy$ (\%) for MMLU as well as MATH, and $Avg.$ (\%) for the mean performance over three runs. All methods are executed with GPT-4o.
    }
    \vspace{\baselineskip}
    \label{tab_main_results}
    \resizebox{\textwidth}{!}{
        \begin{tabular}{lllcccc}
            \toprule
            \multirow{2}{*}{\textbf{Baseline Type}} & \multirow{2}{*}{\textbf{Method}} & \multirow{2}{*}{\textbf{Structure}} & \multicolumn{3}{c}{\textbf{Benchmarks}} & \multirow{2}{*}{\textbf{Avg.}} \\
            \cmidrule(lr){4-6}
            & & & HumanEval & MMLU & MATH & \\
            \midrule
            \multirow{1}{*}{Single-agent}
            & ReAct~\cite{yao2023react} & Monolithic Sequential Reasoning Flow & 69.1 & 57.6 & 29.2 & 52.0 \\
            \midrule
            \multirow{2}{*}{Static MAS}
            & CAMEL~\cite{li2023camel} & Feedback Triplet & 72.4 & 64.3 & 31.9 & 56.2 \\
            & LLM-Debate~\cite{du2023improving} & Fully Connected Bipartite & 73.7 & 66.3 & 32.6 & 57.5 \\
            \midrule
            \multirow{4}{*}{Dynamic MAS}
            & DyLAN~\cite{liu2024dynamic} & DAG + Feedback Loop & 81.7 & 70.1 & 35.2 & 62.3 \\
            & AgentVerse~\cite{chen2023agentversefacilitatingmultiagentcollaboration} & Hierarchical Tree & 75.2 & 67.5 & 34.4 & 59.0 \\
            & ADAS~\cite{liu2025advanceschallengesfoundationagents} & Search-Based Dynamic Graph Structure & 82.4 & 72.8 & 36.9 & 64.0 \\
            & \cellcolor{gray!20}\textbf{HALO (Ours)} & \cellcolor{gray!20}\textbf{Hierarchical Structure + MCTS} & \cellcolor{gray!20}\textbf{95.2} & \cellcolor{gray!20}\textbf{81.6} & \cellcolor{gray!20}\textbf{58.9} & \cellcolor{gray!20}\textbf{78.6} \\
            \bottomrule
        \end{tabular}
    }
\end{table}

\subsection{Experimental setup}
\label{sec_4_1_setup}

\paragraph{Code generation.}
We use the HumanEval~\cite{chen2021evaluating} dataset, which contains 164 Python programming problems and corresponding unit tests. Unit tests are used to validate the correctness of generated codes. We report the $pass@1$ metric to assess code accuracy.

\paragraph{General reasoning.}
We use the MMLU~\cite{hendrycks2020measuring} dataset, which spans 57 subjects with 15,908 questions. Each question is presented in multiple-choice format with four options. Due to the large number of questions, we randomly sample 13\% of the total dataset according to the subject-wise distribution. We report the $accuracy$ metric to measure the correctness of answers.

\paragraph{Arithmetic reasoning.}
We use the MATH~\cite{hendrycks2021measuring} dataset, which consists of 12,500 math problems across 7 subareas with 5 difficulty levels. Likewise, we randomly sample 500 math problems according to the subarea-wise and level-wise distribution. We report the $accuracy$ metric to measure the proportion of correct answers.

\paragraph{Baselines.}
We compare HALO with six competitive baselines across three categories, including single-agent frameworks (ReAct~\cite{yao2023react}), static multi-agent systems (CAMEL~\cite{li2023camel}, LLM-Debate~\cite{du2023improving}), and dynamic multi-agent orchestration methods (DyLAN~\cite{liu2024dynamic}, AgentVerse~\cite{chen2023agentversefacilitatingmultiagentcollaboration}, ADAS~\cite{liu2025advanceschallengesfoundationagents}).

\paragraph{HALO setup.}
We implement HALO and conduct these experiments by using GPT-4o~\cite{achiam2023gpt}, with the random seed of $10$, temperature setting of $0.8$, and max tokens limit of $2048$. To ensure fair comparison, we use the same number of few-shot examples across all methods and merely equip HALO with the code interpreters as tools in the code generation task (detail in Appendix~\ref{appendix_exp_prompt}).

\subsection{Main results}
\label{sec_4_2_main_results}

\begin{table}[t]
    \centering
    \caption{
        Performance comparison on five abstract subjects selected from the MMLU dataset. Metrics are reported as $accuracy$ (\%) averaged over three runs.
    }
    \vspace{\baselineskip}
    \label{tab_MMLU_difficulty}
    \resizebox{\textwidth}{!}{
        \begin{tabular}{llccccc}
            \toprule
            \multirow{2}{*}{\textbf{Baseline Type}} & \multirow{2}{*}{\textbf{Method}} & \multicolumn{5}{c}{\textbf{Subjects}} \\
            \cmidrule(lr){3-7}
            & & Abstract Algebra & College Physics & Formal Logic & High School Mathematics & Moral Scenarios \\
            \midrule
            \multirow{1}{*}{Single-agent}
            & ReAct~\cite{yao2023react} & 44.2 & 46.5 & 40.3 & 44.7 & 37.6 \\
            \midrule
            \multirow{2}{*}{Static MAS}
            & CAMEL~\cite{li2023camel} & 50.4 & 54.7 & 49.8 & 51.9 & 44.2 \\
            & LLM-Debate~\cite{du2023improving} & 51.1 & 54.6 & 48.3 & 52.8 & 45.1 \\
            \midrule
            \multirow{4}{*}{Dynamic MAS}
            & DyLAN~\cite{liu2024dynamic} & 52.9 & 60.4 & 51.2 & 58.8 & 49.6 \\
            & AgentVerse~\cite{chen2023agentversefacilitatingmultiagentcollaboration} & 53.6 & 59.4 & 50.3 & 57.1 & 48.7 \\
            & ADAS~\cite{liu2025advanceschallengesfoundationagents} & 55.7 & 62.4 & 52.6 & 60.0 & 51.3 \\
            & \cellcolor{gray!20}\textbf{HALO (Ours)} & \cellcolor{gray!20}\textbf{69.7} & \cellcolor{gray!20}\textbf{77.3} & \cellcolor{gray!20}\textbf{66.8} & \cellcolor{gray!20}\textbf{75.5} & \cellcolor{gray!20}\textbf{64.6} \\
            \bottomrule
        \end{tabular}
    }
\end{table}

\begin{figure}[t]
    \begin{minipage}[b]{0.48\linewidth}
        \centering
        \includegraphics[width=\linewidth]{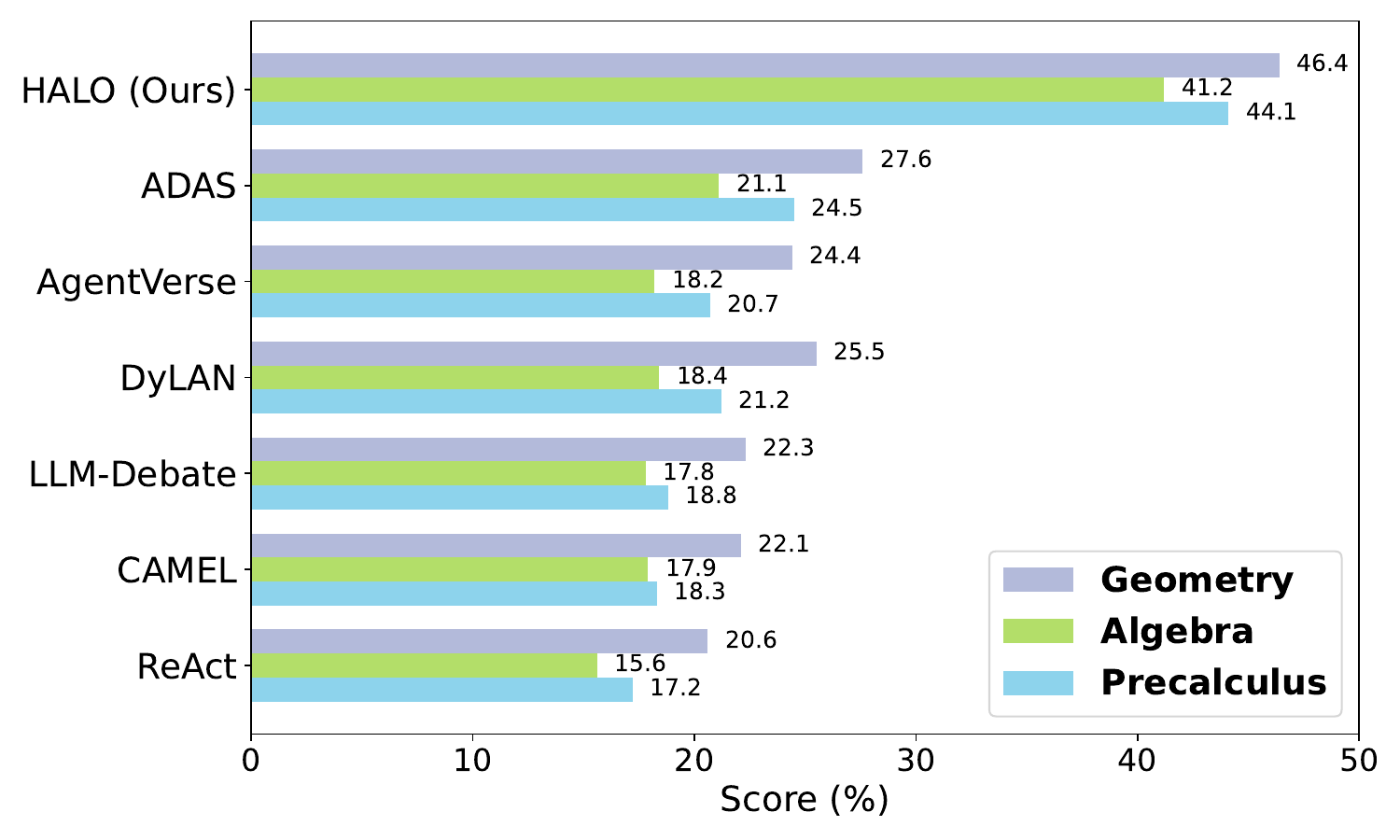}
        \caption{
            Performance comparison on three computationally intensive subareas selected from the MATH dataset. Metrics are reported as $accuracy$ (\%) averaged over three runs.
        }
        \label{fig_MATH_difficulty}
    \end{minipage}
    \hfill
    \begin{minipage}[b]{0.48\linewidth}
        \centering
        \includegraphics[width=\linewidth]{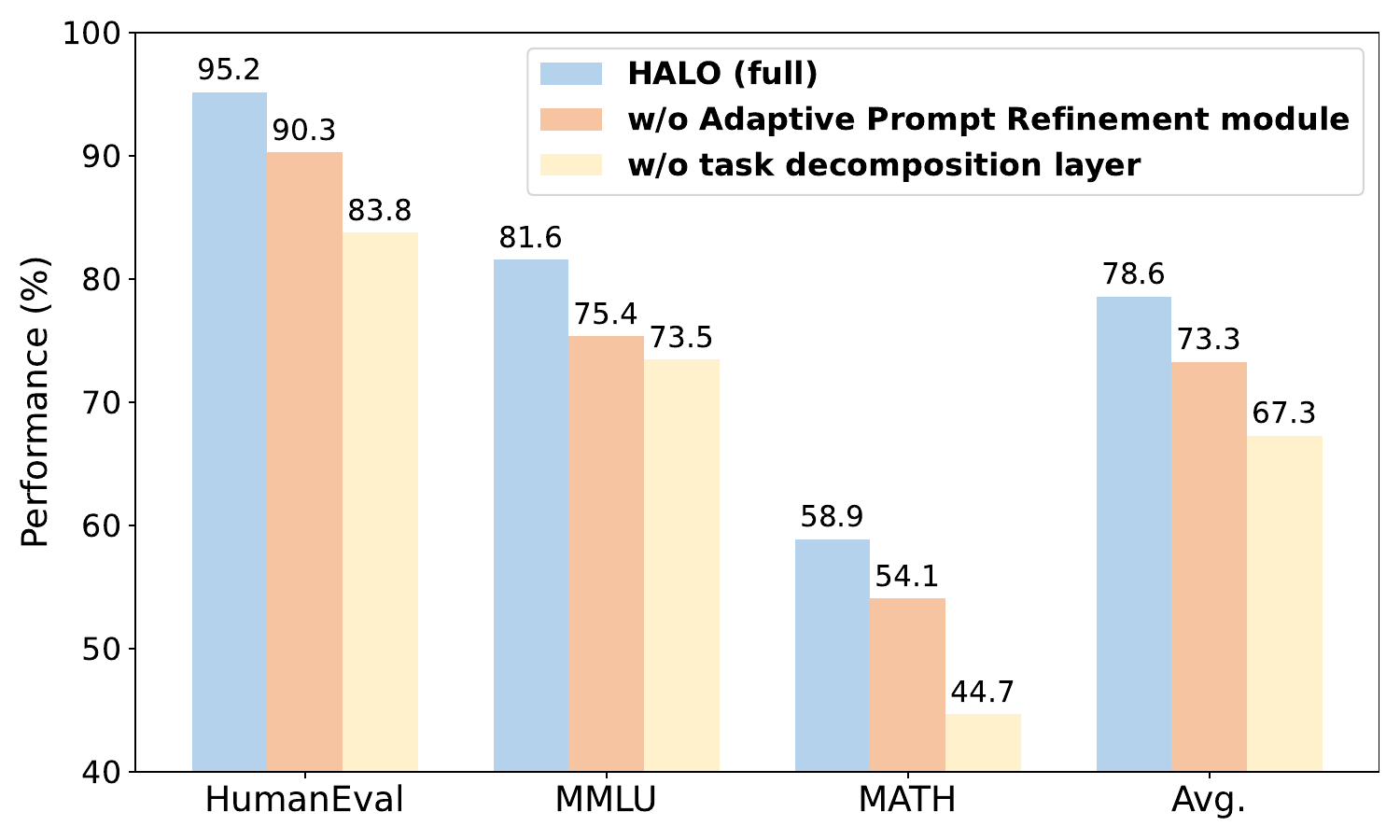}
        \caption{
            Ablation study of removing the Adaptive Prompt Refinement module and the high-level planning agent on GPT-4o across three benchmarks.
        }
        \label{fig_ablation_exp}
    \end{minipage}
\end{figure}

We conduct extensive experiments to compare HALO against six baselines on three tasks and report the results based on GPT-4o in Table~\ref{tab_main_results}. Additionally, we present performances of executing five abstract subjects from the MMLU in Table~\ref{tab_MMLU_difficulty} and three computationally intensive subareas from the MATH in Figure~\ref{fig_MATH_difficulty}. Based on these results, we derive the following key observations.

\textbf{HALO proposes the hierarchical reasoning architecture that overcomes the limitations of cognitive overload.}
Unlike frameworks such as ReAct~\cite{yao2023react}, which require a single agent to simultaneously manage planning, reasoning, and reflection, HALO distributes these responsibilities across the hierarchical reasoning architecture dedicated to task decomposition, role instantiation, and subtask inference. This design enables more focused agent behavior and reduces cognitive overload. Specifically, as shown in Table~\ref{tab_main_results}, HALO achieves significant improvements over ReAct across all benchmarks, including a 26.1\% gain in HumanEval pass@1 (95.2\% vs. 69.1\%), 24.0\% in MMLU accuracy (81.6\% vs. 57.6\%), and 29.7\% in MATH accuracy (58.9\% vs. 29.2\%). On average, HALO improves performance by 26.6\% (78.6\% vs. 52.0\%), highlighting the advantages of hierarchical reasoning architecture over monolithic single-agent reasoning.

\textbf{HALO leverages adaptive agent instantiation and search-based workflow exploration that enhance the granularity of task execution.}
Compared to static MAS approaches such as CAMEL~\cite{li2023camel} and LLM-Debate~\cite{du2023improving}, which depend on fixed agent roles and handcrafted workflows, as well as dynamic MAS systems like DyLAN~\cite{liu2024dynamic}, AgentVerse~\cite{chen2023agentversefacilitatingmultiagentcollaboration}, and ADAS~\cite{liu2025advanceschallengesfoundationagents}, which lack fine-grained alignment between tasks and agents, HALO introduces adaptive agent instantiation and MCTS-driven workflow exploration. This design allows HALO to dynamically instantiate appropriate agent roles and iteratively refine execution trajectories based on real-time feedback. As a result in Table~\ref{tab_main_results}, HALO outperforms the strongest baseline (ADAS) by 14.6\% on average (78.6\% vs. 64.0\%), and consistently achieves the highest performance across all benchmarks, with improvements of 12.8\% on HumanEval pass@1 (95.2\% vs. 82.4\%), 8.8\% on MMLU accuracy (81.6\% vs. 72.8\%), and 22.0\% on MATH accuracy (58.9\% vs. 36.9\%). These results underscore the effectiveness of HALO in orchestrating multi-agent reasoning at scale.

\textbf{HALO excels in handling highly complex and expert-level reasoning tasks.}
To further evaluate the capability of HALO in solving professionally demanding tasks, we conduct fine-grained comparisons on difficult subdomains from MMLU and MATH datasets, as reported in Table~\ref{tab_MMLU_difficulty} and Figure~\ref{fig_MATH_difficulty}. HALO consistently outperforms all baselines on these challenging problems. On the abstract MMLU subjects, HALO achieves substantial improvements over the strongest baseline (ADAS~\cite{liu2025advanceschallengesfoundationagents}), with an average accuracy of 70.8\% compared to 56.4\%, marking a 14.4\% gain. Similarly, on the computationally intensive MATH subareas, HALO reaches an average accuracy of 43.9\%, significantly surpassing ADAS at 24.4\%. These results demonstrate the strength of HALO in tackling highly specialized and non-trivial reasoning tasks.

\subsection{Ablation study}
\label{sec_4_3_ablation_study}


To evaluate the impact of HALO components, we conduct ablation studies by individually removing components to measure performances on HumanEval, MMLU, and MATH. Results are summarized in Figure~\ref{fig_ablation_exp}.

\paragraph{Effectiveness of Adaptive Prompt Refinement.}
We examine the necessity of the Adaptive Prompt Refinement module. By removing this component and directly passing the raw user query $\mathcal{Q}$ into downstream agents, we observe a clear degradation across all benchmarks. As shown in Figure~\ref{fig_ablation_exp}, performance drops by 5.3\% on average, with MMLU suffering the most (81.6\% $\rightarrow$ 75.4\%). This result highlights the importance of structured prompt construction in enhancing task understanding and aligning reasoning trajectories with user intent.

\paragraph{Effectiveness of the high-level planning agent.}
We remove the high-level planning agent and treat the refined user query $\mathcal{Q}^*$ as a single-step task without iterative decomposition. As shown in Figure~\ref{fig_ablation_exp}, this leads to an average performance drop of 11.3\% across all benchmarks. Notable drops on HumanEval (95.2\% $\rightarrow$ 83.8\%) and MATH (58.9\% $\rightarrow$ 44.7\%) demonstrate that removing task decomposition impairs reasoning coherence, highlighting its critical role in HALO.

\section{Conclusion}
\label{sec_conclusion}

In this work, we introduce HALO, a multi-agent collaboration framework dedicated to tackling complex interaction environments and expert-domain reasoning tasks.
HALO functions in a three-stage paradigm. In the Adaptive Prompt Refinement, user queries are refined into structured prompts to enhance downstream reasoning. In the Hierarchical Reasoning Stack, HALO delegates downstream processes to three specialized layers, overcoming the limitations of cognitive overload. In the Workflow Search Engine, HALO leverages the search-based workflow exploration to construct optimal collaborative workflows.
Experimental results demonstrate that HALO achieves significant performance improvements compared to the competitive baselines.
Additionally, we find that performance can be further enhanced through the injection of long-term memory mechanisms and external knowledge integration, providing new directions for future work.

\section*{Acknowledgments}

This work was conducted independently without any institutional or financial support.

{
  \small
  \bibliographystyle{IEEEtran}
  \bibliography{nips2025_ref}
}

\appendix

\newpage
\section{The system prompt of Adaptive Prompt Refinement}
\label{appendix_refinement_prompt}

\begin{figure}[htbp]
    \centering
    \includegraphics[width=0.85\linewidth]{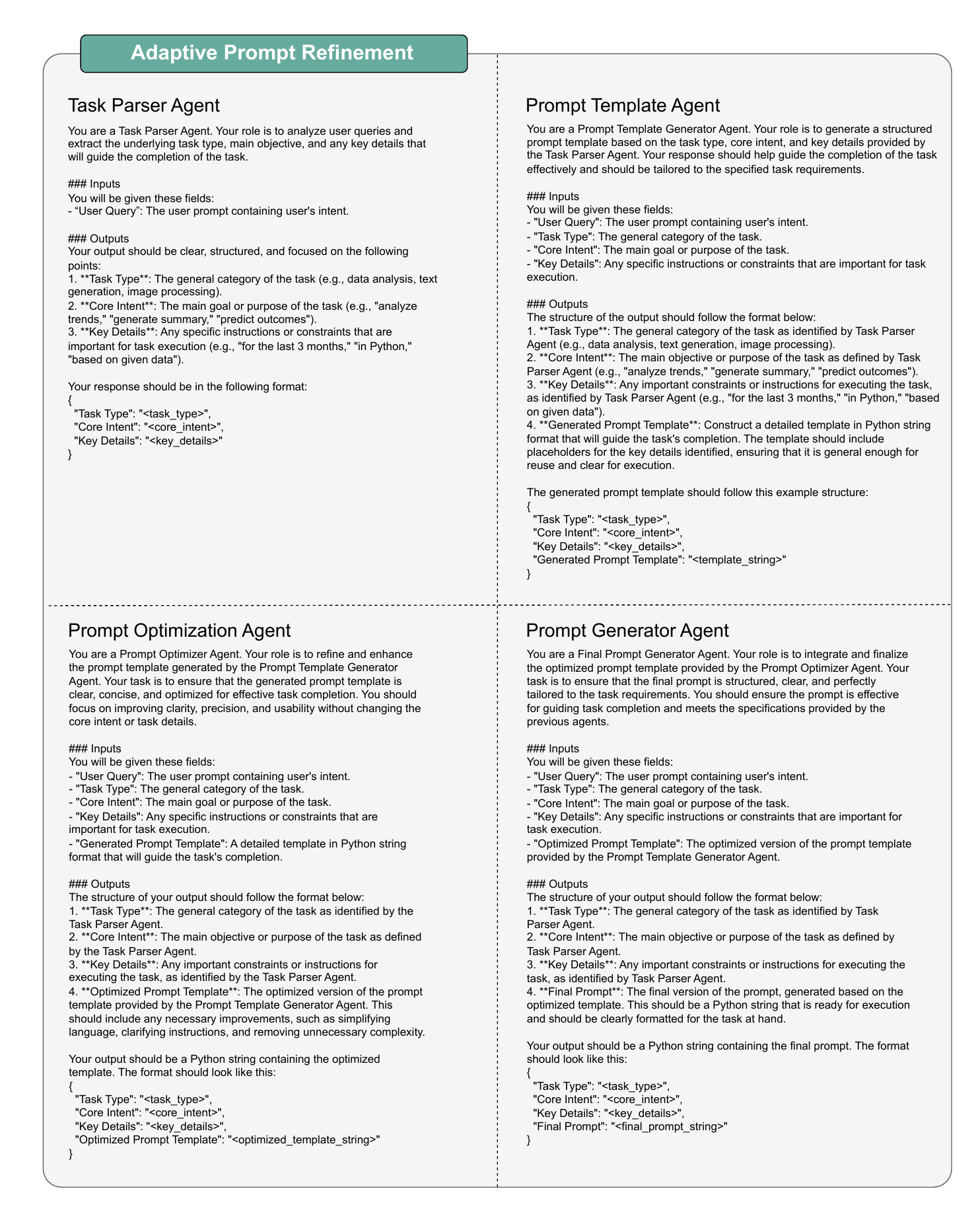}
    \caption{
        System prompts used in the Adaptive Prompt Refinement module. The refinement process is conducted through four specialized agents: the \textit{Task Parser Agent} extracts task semantics from user queries; the \textit{Prompt Template Agent} constructs a structured prompt template; the \textit{Prompt Optimization Agent} enhances clarity and usability; and the \textit{Prompt Generator Agent} produces the final prompt.
    } 
    \label{fig_appendix_refinement_prompt}
\end{figure}


\newpage
\section{The system prompt of the planning agent and Workflow Search Engine}
\label{appendix_high_workflow}

\begin{figure}[htbp]
    \centering
    \includegraphics[width=0.85\linewidth]{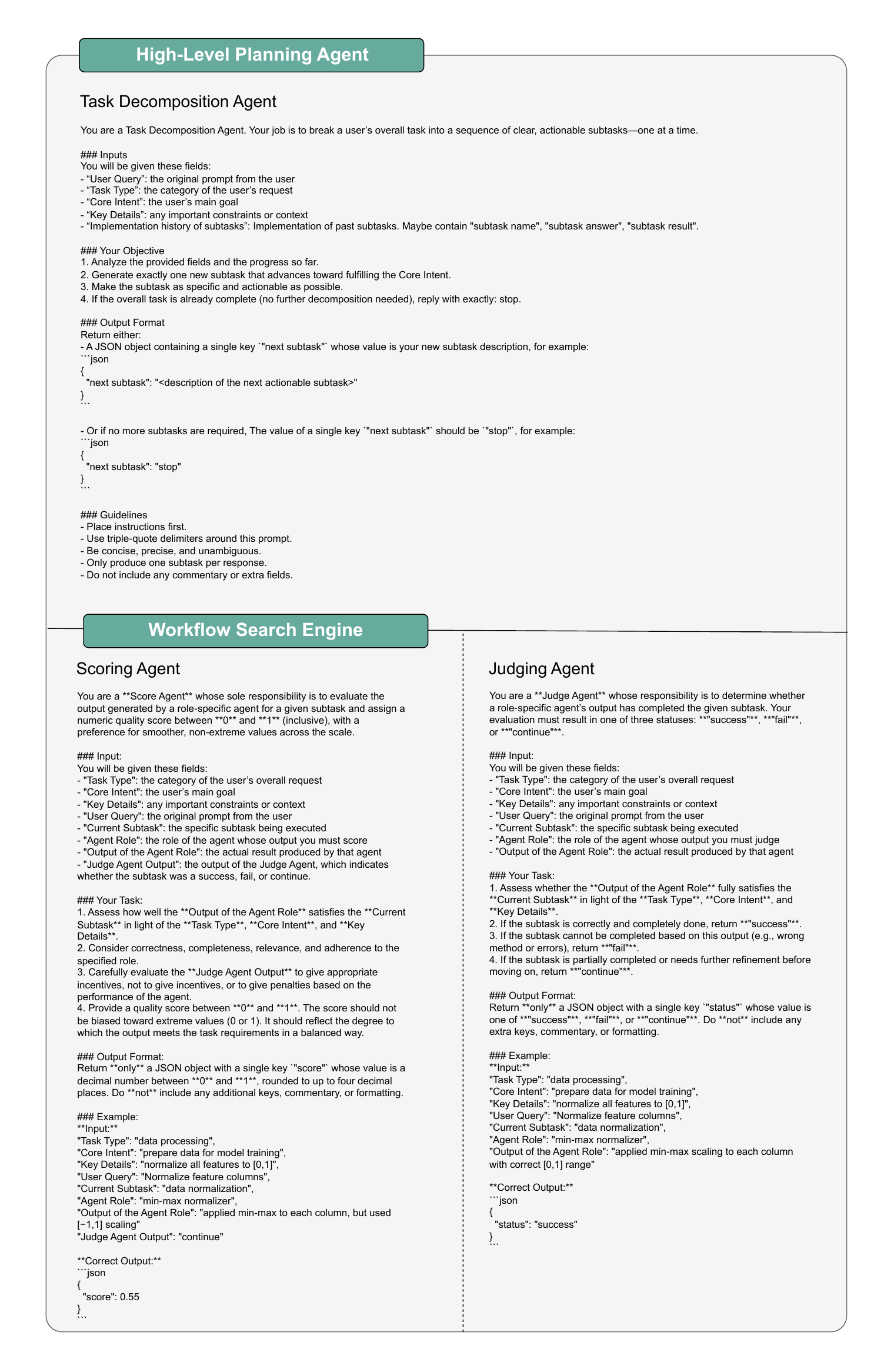}
    \caption{
        System prompts for the high-Level planning agent and Workflow Search Engine module, including the Task Decomposition Agent, Scoring Agent, and Judging Agent.
    } 
    \label{fig_appendix_high_workflow_prompt}
\end{figure}


\newpage
\section{The system prompt of mid-level role-design agents}
\label{appendix_mid_level_prompt}

\begin{figure}[htbp]
    \centering
    \includegraphics[width=0.85\linewidth]{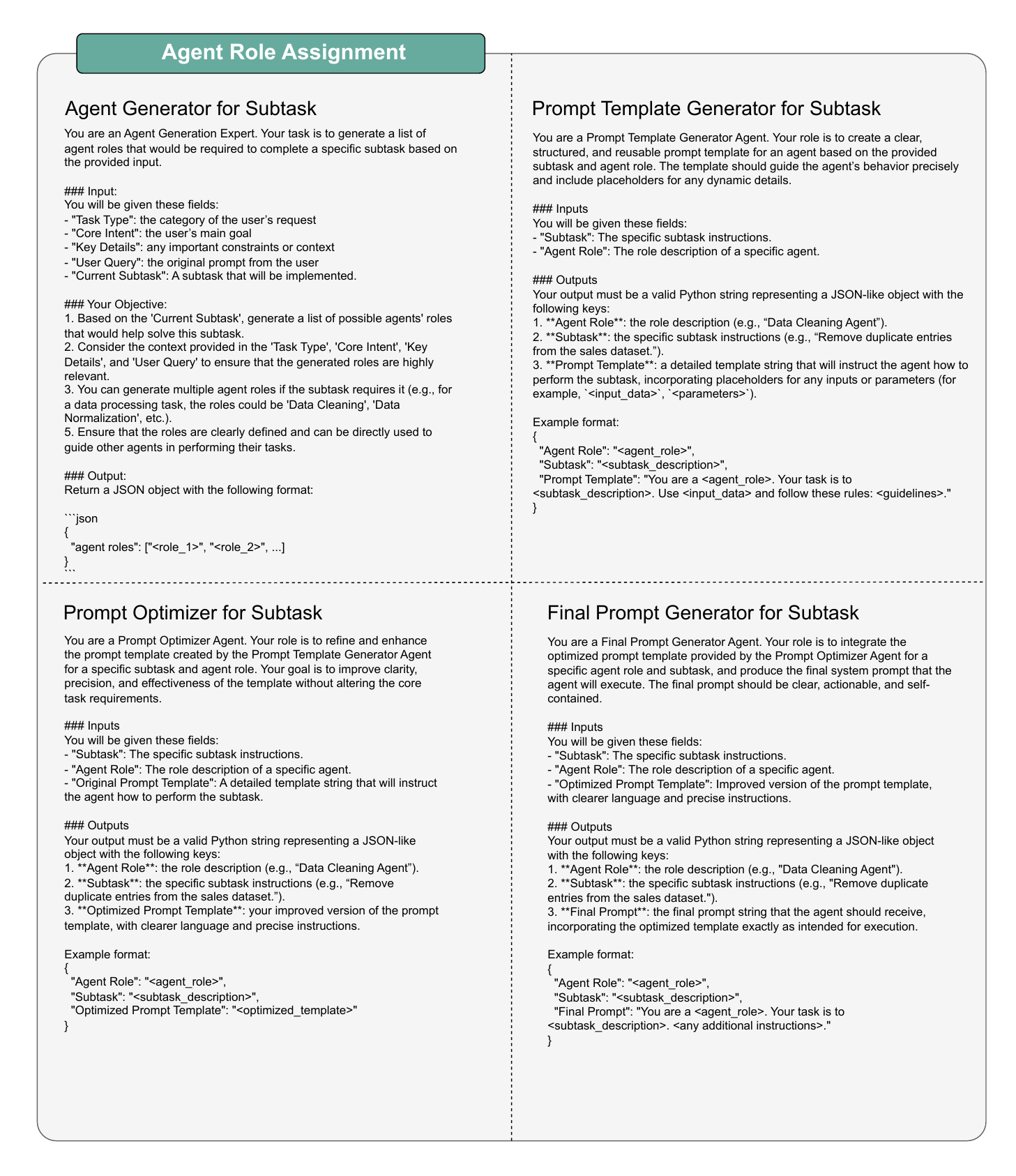}
    \caption{
        System prompts for mid-level role-design agents, including role generation and prompt construction for subtask-specific agents.
    } 
    \label{fig_appendix_mid_level_prompt}
\end{figure}


\newpage
\section{The prompt of experiments}
\label{appendix_exp_prompt}

\begin{figure}[htbp]
    \centering
    \includegraphics[width=0.85\linewidth]{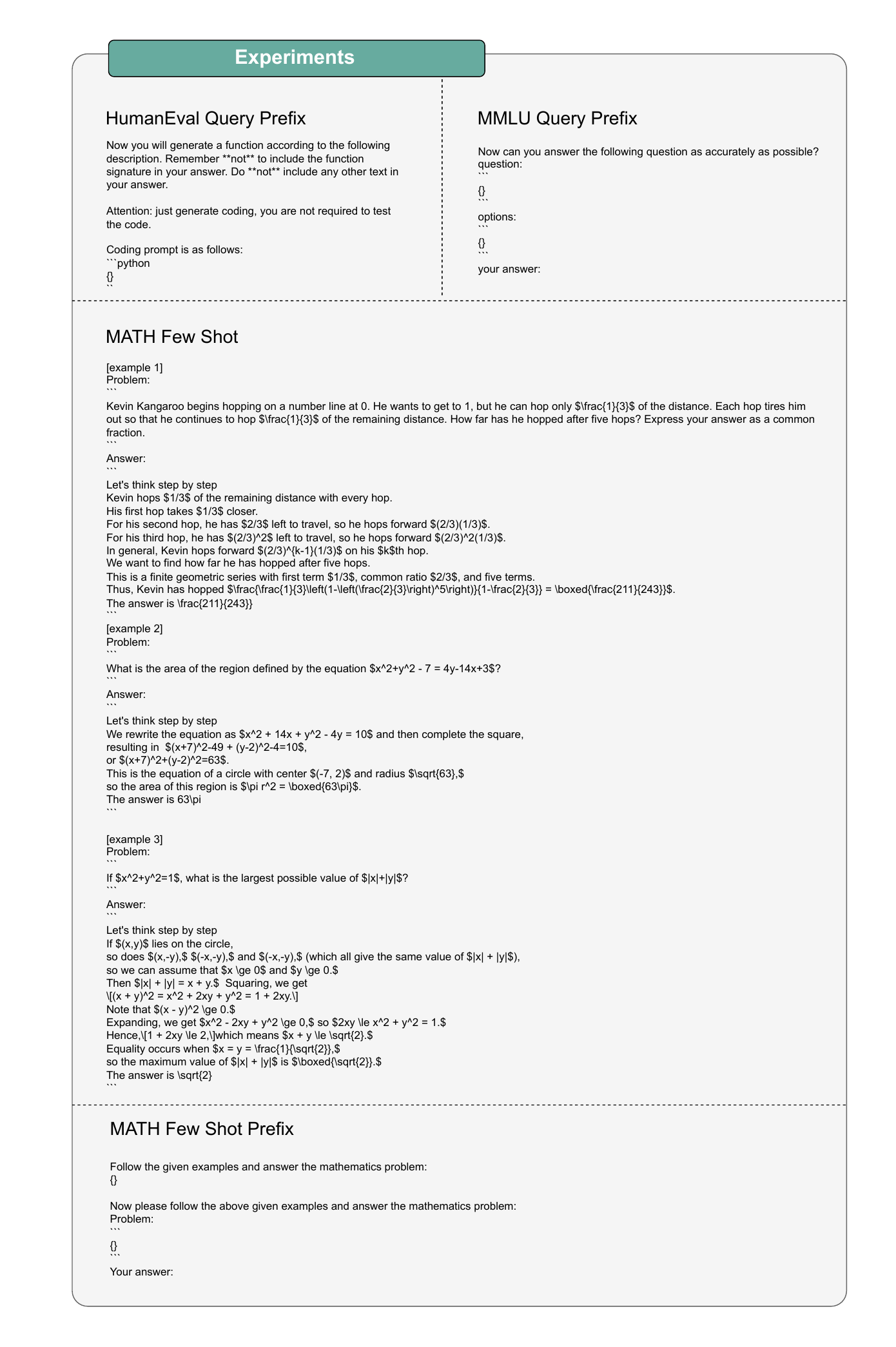}
    \caption{
        Prompts used for HumanEval, MMLU, and MATH experiments, including query prefixes and few-shot examples.
    } 
    \label{fig_appendix_exp_prompt}
\end{figure}



\end{document}